\documentstyle[12pt,epsfig,rotating]{article}
\begin{document}

December 1994  \hfill  DESY 94--247

\hfill hep-ph/9501270

\vspace{2cm}

\begin{center}

{\Large \bf{Single Spin Asymmetries in Proton-Proton}}   \\ [5pt]
{\Large \bf{and Proton-Neutron Scattering at 820 GeV} \footnote{Updated
version of a talk given at SPIN '94 conference,
Bloomington, Indiana, Sept. 15-22, 1994 }}   \\ [40pt]

{\large Wolf-Dieter Nowak~\footnote{e-mail: nowakw@ifh.de}}    \\ [30pt]

{\small \it DESY-Institut f\"ur Hochenergiephysik \\  [5pt]
Platanenallee 6   \\
 D-15738 Zeuthen, Germany }             \\ [3cm]

{\bf Abstract}
\end{center}
\smallskip

    The physics case is summarised for the investigation of high energy spin
    pheno\-mena by placing an internal polarised target into HERA's unpolarised
    proton beam. The luminosity and experimental sensitivity are discussed.
    Estimating the physics reach of single spin asymmetries in different
    final states reveals a consi\-de\-rable physics potential in
    testing the spin sector of perturbative QCD.

\newpage

\section{Introduction}

If the HERA proton ring could be operated with polarised particles a host of
new experimental possibilities would ensue. At present it is not clear though
whether this is a practical possibility. A viable first step, however, towards
investigating spin effects in hadron--hadron interactions at HERA could be
an expe\-riment scattering unpolarised 820 GeV protons off polarised
nucleons uti\-li\-sing a polarised internal gas target. Both proton-proton
and proton-neutron spin asymmetries would be readily accessible
since modern polarised gas targets can be operated with Hydrogen,
Deuterium,  or $^3$He.

Experimentally single spin asymmetries are usually generated by swit\-ching
the direction of the initial spin vector. In the given situation they can
be measured as correlation between the polarisation of the target nucleon
on one hand, and the final state polarisations and angular distributions
on the other.

In the next section the experimental sensitivity is
estimated for two different luminosity scenarios. Then, the physics reach with
several interesting final states is discussed in some detail. The magnitude
of the gluon spin might be probed with jets or possibly dimuons. It appears
feasible to measure (leading) twist-3 contributions with direct photons
where theoretical predictions distinguishing between proton and neutron
target are of special interest. New pion data at higher transverse
momenta could clarify the situation of the non-zero transverse pion asymmetry
which appears to be in conflict with pertubative QCD. Analysing dipions from
the same jet might allow to access the valence quark transversity
distribution.  \\

\bigskip
\section{Luminosity and Sensitivity}        \label{sect_lumi}

A realistic estimate for the average HERA proton beam current is
$\bar{I}_B = 80 \; \mbox{mA} = 0.5 \cdot 10^{18} \; s^{-1}$
constituting half of the design current.
An internal target area desity of $n_T = 10^{12}$ atoms/cm$^2$, as can
be delivered by a standard polarised jet target, would not deteriorate
the present HERA proton beam performance. This
safe ''low luminosity option'' would have a luminosity of
\begin{eqnarray*}
{\cal{L}}_L = n_T \cdot \bar{I}_B = 0.5 \cdot 10^{30} \; \mbox{cm}^{-2}
s^{-1} \; .
\end{eqnarray*}

\noindent In an optimistic scenario the polarised internal target could
presumably be operated with  an area density of a few $10^{13}$, say
$n_T = 3 \cdot 10^{13}$ atoms/cm$^2$. Note that
the UA6 unpolarised internal target was successfully run at
a comparable density in the CERN $Sp\bar{p}S$ collider. Today's
polarised H/D \cite{zap1} and $^3$He targets \cite{kra1} with storage
cells are capable of running at those densities with polarisations as
high as 80\% and 50\%, respectively. Hence a ''high luminosity
option'' with
\begin{eqnarray*}
{\cal{L}}_H = 1.5 \cdot 10^{31} \; \mbox{cm}^{-2} s^{-1}
\end{eqnarray*}

\noindent appears feasible although still to be proven under
actual HERA conditions.   \\

To assess the physics reach of different final states a total running
time of $T = 1.6 \cdot 10^7 \; s$ with 100\% efficiency is assumed.
This corresponds to about 3 calendar years of HERA operation with 6 months per
year physics running and 33\% combined up--time for accelerator and experiment.
Hence the integrated luminosities per year for the two discussed running
scenarios are \\

\begin{center}
\begin{tabular}{lp{6cm}}
${\cal{L}}_L \cdot T = \, \, 8 \; pb^{-1}$  & for the low luminosity option
and \\
\vspace{5pt}
${\cal{L}}_H \cdot T = 240 \; pb^{-1}$  & for the high luminosity option.
\end{tabular}
\end{center}
The experimental sensitivity in the measured single spin asymmetry A is

\vspace{-10pt}
\begin{eqnarray*}
\delta A = \frac{1}{p_{targ}} \cdot \frac{1}{\sqrt{N}} \; ,
\end{eqnarray*}

\noindent
where $p_{targ}$ is the degree of target polarisation and
$N = {\cal{L}} \cdot T \cdot C \cdot \sigma  \;$
the total number of recorded events. Here $\sigma$ is the
unpolarised cross section and C the combined trigger and reconstruction
efficiency. Then

\vspace{-10pt}
\begin{eqnarray*}
\delta A = \frac{1}{p_{targ}} \cdot \frac{1}{\sqrt{{\cal{L}} \cdot T
    \cdot C}} \cdot \frac{1}{\sqrt{\sigma}} \; ,
\end{eqnarray*}

\noindent
and with $p_{targ} = 0.8$ and C = 50\% one obtains as
experimental sensitivities

\begin{center}
\begin{tabular}{lp{6cm}}
$\delta A_L = 0.6 / \sqrt{\sigma \, [pb]}$ &  for the low
luminosity option and \\
\vspace{5pt}
 $\delta A_H = 0.1 / \sqrt{\sigma \, [pb]}$ & for the high
luminosity option.
\end{tabular}
\end{center}

\bigskip
\section{Physics Objectives}   \label{sect_phys}

\subsection{Probing the Gluon Spin with Inclusive Jets}

At $\sqrt{s}$ = 40 GeV
and $p_t$~=~5 GeV the huge unpolarised cross section for inclusive jet
production allows for sensitivities of 0.0001 [0.0006]
in the high [low] luminosity option. At $p_t$~=~10 GeV the sensitivities are
still 0.003 [0.018], always meant for 1 GeV bins. Stratmann and Vogelsang
\cite{str1} calculated the corresponding hard scattering cross sections for
both {\it transverse} and {\it longitudinal} singly polarised
proton--proton scattering including all underlying pQCD 2~$\rightarrow$~2
subprocesses. The {\it transverse} asymmetry is shown in fig. 1, at
$p_t$~=~10 GeV ($x_t$~=~0.5) is $A_T \simeq 12\%$, unfortunately with
insignificant dependence on the transverse  polarisation of the sea.

%+++++++++++ Hier FIG.1 plazieren. +++++++++++++
\setlength{\unitlength}{1mm}
\begin{figure}[t]
\begin{center}
\begin{picture}(160,80)(0,0)
\begin{turn}{90}%
%\special{psfile=bild1.ps voffset=-50 hoffset=-320 hscale=70
%  vscale=70 angle=0}
\end{turn}
\end{picture}
\end{center}
{\small {\sl Fig. 1} {\em Transverse asymmetry on the parton level
for inclusive jet production at
$\sqrt{s}$ = 40 GeV. The solid line was obtained using the scale
$Q^2 = p_t^2/4$, the dashed one corresponds to the fixed scale
$Q^2$ = 10 GeV$^2$.}}

\end{figure}

The {\it longitudinal} asymmetry $A_L$ is very sensitive to the size of
$\Delta$G, as can be seen from fig. 2. Over the accessible $p_t$ range
5~$\div$~10 GeV $A_L$ rises smoothly from 5 to 25\% if $\Delta$G~=~0,
whereas it stays approximately constant at 25\% when a very
large gluon spin is assumed.

%+++++++++++++ Hier FIG.2 plazieren. +++++++++++++
\setlength{\unitlength}{1mm}
\begin{figure}[t]
\begin{center}
\begin{picture}(160,80)(0,0)
\begin{turn}{90}%
%\special{psfile=bild2.ps voffset=-50 hoffset=-330 hscale=70
%  vscale=70 angle=0}
\end{turn}
\end{picture}
\end{center}
{\small {\sl Fig. 2} {\em Longitudinal asymmetry on the parton level
for inclusive jet production
at $\sqrt{s}$ = 40 GeV \cite{str2}, based upon
two different sets of polarized parton distributions. Set a) implies a very
large gluon spin and a vanishing polarized sea, set b) has a vanishing gluon
spin but a large negative polarized sea, cf. \cite{str1}.}}
\end{figure}

In the given kinematical situation the c.m. backward jet will emerge under
a few hundred milliradian in the laboratory system. Anticipating that
the three fastest particles in the jet can be isolated a handedness
\cite{efr1,bel1}
analysis is believed to measure the spin of the fragmenting parton. The
supposedly process independent handedness parameter could possibly be
about 0.05 \cite{efr2}. This, together with dilutions from the internal target
polarisation (0.8) and an anticipated average parton spin (0.25),
would result in a total dilution factor of about 100. Hence the 25\%
parton level asymmetry would be reduced to a 2.5~per mille, i.e. 0.0025
hadronic level asymmetry. This, at $p_t$~=~5~GeV, is a 4~$\sigma$ effect
even in the low luminosity option. Obviously, the systematic error has to
be kept on the permille level as well.

\medskip
\subsection{Probing the Gluon Spin with Dimuons}

Carlitz and Willey \cite{car1}
calculated the {\it longitudinal} single spin asymmetry $A_L$ for dimuon
production in proton-proton collisions. It is non-zero if the axial vector
built from the muon momenta has a longitudinal component. The relevant
subprocesses are gluon Compton scattering and quark--antiquark annihilation.
Using different assumptions on the total gluon spin this asymmetry was
calculated by Nadolsky \cite{nad1}. As shown in fig. 3, at
$Q^2_{\perp}~=~3~(GeV)^2$
the asymmetry ranges from $A_L$~=~0.02 for $\Delta$G~=~1  up to
$A_L~=~0.08~\div~0.12$ for $\Delta$G~=~6, although today $\Delta$G~=~3~$\div$~4
might be more a realistic upper limit.
For dimuon masses above 10 GeV this scenario corresponds to an unpolarised
cross section below 1 pb. Even in the high luminosity option the expected
experimental sensitivity is at best $\delta A \simeq 0.1$ and hence
comparable to the whole asymmetry difference.

%+++++++++++++ Hier FIG.3 plazieren. ++++++++++++++
\setlength{\unitlength}{1mm}
\begin{figure}[t]
\begin{center}
\begin{picture}(160,80)(0,0)
\begin{turn}{90}%
%\special{psfile=bild3.ps voffset=-50 hoffset=-320 hscale=70
%  vscale=70 angle=0}
\end{turn}
\end{picture}
\end{center}
{\small {\sl Fig. 3} {\em Longitudinal asymmetry on the parton level
for dimuon production at 400
GeV incoming energy \cite{nad2} versus virtual photon energy for different
assumptions on the gluon spin. At 820 GeV the asymmetries will be slightly
smaller.}}
\end{figure}

The steeply rising dimuon cross section suggests to reconsider the
case at smaller dimuon masses, e.g. 4 GeV. Here the cross section of
about 100 pb leads to $\delta A \simeq 0.01$ in the high luminosity option.
However, the actual improvement in the sensitivity to asymmetry ratio remains
questionable since in most models a smaller gluon spin is expected when
it is probed at smaller dimuon masses.

\medskip
\subsection{Probing Twist-3 Matrix Elements with Direct Photons}

Based upon
a twist-3 parton distribution involving the correlation between quark
fields and the gluonic field strength, the leading single {\it transverse}
spin asymmetry for high $p_{t}$ direct photon production
was estimated by Qiu and Sterman \cite{qiu1}. The essential subprocess
is gluon Compton scattering with the gluon carrying the initial polarisation
information. The hard scattering asymmetry rises to about 20\% for
$x_{F} \simeq$ -~0.8.
Estimating the same matrix element differently Ehrnsperger et al. \cite{ehr1}
reconsidered the case with special emphasis to differences between proton
and neutron target. Only a small negative proton asymmetry but a rather
large positive neutron asymmetry of several 10\% is predicted.
According to Korotkiyan and Teryaev \cite{kor1} the contribution due to
gluonic poles should vanish and all asymmetry is due to fermionic poles,only.
Nevertheless, a sizeable parton level asymmetry remains, as can be seen from
fig. 4.

%+++++++++++++ Hier FIG.4 plazieren. +++++++++++++++
\setlength{\unitlength}{1mm}
\begin{figure}[t]
\begin{center}
\begin{picture}(160,80)(0,0)
\begin{turn}{90}%
%\special{psfile=bild4.ps voffset=-50 hoffset=-320 hscale=70
%  vscale=70 angle=0}
\end{turn}
\end{picture}
\end{center}
{\small {\sl Fig. 4} {\em Transverse asymmetry on the parton level
for inclusive direct photon production.}}
\end{figure}

In another paper Ji \cite{xji1} considered the variety of relevant
three--gluon--correlations. Many different structure functions appear in the
cross section making any selection impossible when considering the full
kinematic region. At large negative $x_F$ however, as for the
measurements at HERA, the pure gluon correlations are expected to become
dominant compared to quark--gluon correlations.

The dilution of the asymmetry on going from the parton to the
hadron level amounts to a factor of 5,
since there is no fragmentation. Hence the above discussed neutron asymmetry
requires $\delta A \leq 0.005$, i.e. it can be studied up to
$p_t \simeq$~4.5~GeV in the high luminosity option. Note that the
gaseous H/D target considered is a good tool to minimise systematic errors
in the study of proton--neutron differences.    \\

\medskip
\subsection{New Physics from Transverse Pion Asymmetries ?}

The only significant data on single spin asymmetries are from E704 who have
measured the reactions $p^{\uparrow}+ p\rightarrow \pi^{0 \pm} + X$ in a
{\it transversely} polarised beam at 200 GeV \cite{ada1}. In strong
contradiction
to pQCD predictions, significant non--zero asymmetries were found, as is
shown in fig. 5. With increasing $x_F$ the charged pion asymmetry
is smoothly rising to 40\%, opposite in sign for the different pion charges.
Years after publication this data still constitutes a challenge to
perturbative QCD. Confirmation by an independent experiment at twice as
high transverse momenta would certainly be worthwile.

%++++++++++++ Hier FIG.5 plazieren. ++++++++++++++
\setlength{\unitlength}{1mm}
\begin{figure}[t]
\begin{center}
\begin{picture}(160,80)(0,0)
\begin{turn}{90}%
%\special{psfile=bild5.ps voffset=-50 hoffset=-320 hscale=70
%  vscale=70 angle=0}
\end{turn}
\end{picture}
\end{center}
{\small {\sl Fig. 5} {\em Transverse asymmetries measured by E704 in inclusive
pion production at 200 GeV.}}
\end{figure}

An example of heretic ideas on this subject is provided by the model
of orbiting valence quarks proposed by Boros et al. \cite{bor1}. Within a
semi--classical approach the constituents of a polarised hadron are
assumed to perform an orbital motion about the polarisation axis and
left/right asymmetries are expected to arise from annihilations of these
valence quarks. The model is able to give a fair
description of the inclusive pion data and of the dimuon data as well.  \\

In a very recent QCD based approach \cite{ter1} is argued that in high $p_t$
inclusive pion production the gluon--quark asymmetry on the parton level
dominates over the gluon--gluon one. As a consequence the flavor--dependent
valence quark polarization leads to a mirror symmetry
in the sign of charged pion asymmetries and explains qualitatively the
behaviour of the $\pi^0$ asymmetry as well.

\bigskip
\subsection{Accessing the Valence Quark Transversity with Dipions}

The measurement
of  two--pion correlations within the same jet, in a scattering of
{\it transversely} polarised hadrons off unpolarised ones, is proposed
by Collins et al. \cite{col1}. To measure the transversity distribution on the
twist-2 level they propose to jointly probe the
transversity of two quarks participating in the hard scattering.
The underlying pQCD subprocess is quark gluon scattering with the outgoing
polarised quark fragmenting into a jet which is then supposed to carry
spin information from the transversely polarised target valence quark.
It is expected that for low--mass pion pairs the azimuthal dependence of
the two--pion plane about the jet axis reverses sign when the spin of the
incoming hadron is reversed.

The reconstruction of the hard scattering would require to measure c.m.
opposite jets associated with pion pairs. Obviously, in the given fixed target
environment the feasibility of pion identification in the forward jet
deserves further study.
The experimental sensitivity was already discussed above. The spin
transfer in the hard subprocess is possibly large hence
a large asymmetry in the overall process might occur.
More theoretical work would is needed to arrive at numerical predictions.

\newpage
\section{Conclusions}                \label{sect_concl}

There is reason for optimism that
interesting and important, and even completely new fundamental
information on the nucleon spin could be accessed by placing an internal
polarised target into the unpolarised HERA proton beam. A rather
broad physics programme aiming at testing the spin sector of perturbative
QCD and beyond, can be based upon measurements of single spin asymmetries
in several final states. However, for most of the accessible processes
better theoretical predictions are necessary in order to
justify a serious experimental effort.

\bigskip
\section*{Acknowledgements}                \label{sect_acknow}

Many thanks are due to D.Trines and E.Steffens for valuable support on beam
and target issues. Enlightening discussions with S.Manayenkov, T.Meng,
M.Ryskin, A.Sch\"afer, J.Soffer, M.Stratmann, O.Teryaev, and
W.Vogelsang are warmly acknowledged. Special thanks go to A.Sch\"afer
and P.S\"oding for critically reading the manuscript.


\newpage

\begin{thebibliography}{99}

\markright{.}

\bibitem{zap1} K.Zapfe, contribution to this conference
\bibitem{kra1} L.Kramer, contribution to this conference
\bibitem{str1} M.Stratmann, W.Vogelsang, {\it Phys. Lett.} {\bf B295},
277 (1992)
\bibitem{str2} M.Stratmann, W.Vogelsang, priv. comm.
\bibitem{efr1} A.V.Efremov et al., {\it Phys. Lett.} {\bf B284}, 394 (1992)
\bibitem{bel1} S.L.Belostotski et al., {\it Z. Phys.} {\bf C63}, 477 (1994)
\bibitem{efr2} A.V.Efremov, contribution to this conference
\bibitem{car1} R.D.Carlitz, R.S.Willey, {\it Phys. Rev.} {\bf D45}, 2323 (1992)
\bibitem{nad1} P.M.Nadolsky, {\it Z. Phys.} {\bf C62}, 109 (1994)
\bibitem{nad2} P.M.Nadolsky, S.M.Troshin, N.E.Tyurin, {\it Int. J. Mod. Phys.}
{\bf 9}, 2505 (1994)
\bibitem{qiu1} J.Qiu, G.Sterman, {\it Phys. Rev. Lett.} {\bf 67}, 2264 (1991)
\bibitem{ehr1} B.Ehrnsperger et al., {\it Phys. Lett.} {\bf B321}, 121 (1994)
\bibitem{kor1} V.M.Korotkiyan, O.V.Teryaev, {\it JINR preprint}
{\bf E2-93-286} (August 1993)
\bibitem{xji1} X.Ji, {\it Phys. Lett.} {\bf B289}, 137 (1992)
\bibitem{ada1} D.L.Adams et al., {\it Phys. Lett.} {\bf B264}, 462 (1991) \\
\hspace{1cm}   D.L.Adams et al., {\it Z. Phys.} {\bf C56}, 181 (1992)
\bibitem{bor1} C.Boros et al., {\it Phys. Rev. Lett.} {\bf 70}, 1751 (1993)
\bibitem{ter1} O.Teryaev, contribution to this conference
\bibitem{col1} J.C.Collins et al., {\it Nucl. Phys.} {\bf B420}, 565 (1994)

\end{thebibliography}
\end{document}